\documentclass[a4paper]{jpconf}
\usepackage{graphicx}
\usepackage{wrapfig}
\usepackage{amsmath}
\usepackage{amssymb}
\usepackage[utf8]{inputenc}

\begin{document}

\hspace{-5cm}

\title{Charge density wave in hydrogen at high pressure}

\hspace*{-1cm}

\author{Ioan B. Magdău and Graeme J. Ackland}

\address{CSEC, SUPA, School of Physics and Astronomy, University of Edinburgh, UK}

\ead{i.b.magdau@sms.ed.ac.uk, gjackland@ed.ac.uk}

\begin{abstract}
We present extensive molecular dynamics (MD) simulations investigating
numerous candidate crystal structures for hydrogen in conditions
around the present experimental frontier (400GPa). Spontaneous phase 
transitions in the simulations reveal a new structure candidate
comprising twofold coordinated chains of hydrogen atoms.  We explain
the electronic structure of this phase in terms of a charge density
wave and calculate its experimental signature.  In detailed tests of
the accuracy of our calculation, we find that $k$-point sampling is
far more important in MD than in static calculations, because of the
freedom it give the atoms to rearrange themselves optimally for the
given sampling.
\end{abstract}

\section{Introduction}
\noindent
\begin{wrapfigure}{rh}{0.59\textwidth}
\vspace{-75pt}
\begin{center}
\includegraphics[width=0.59\textwidth]{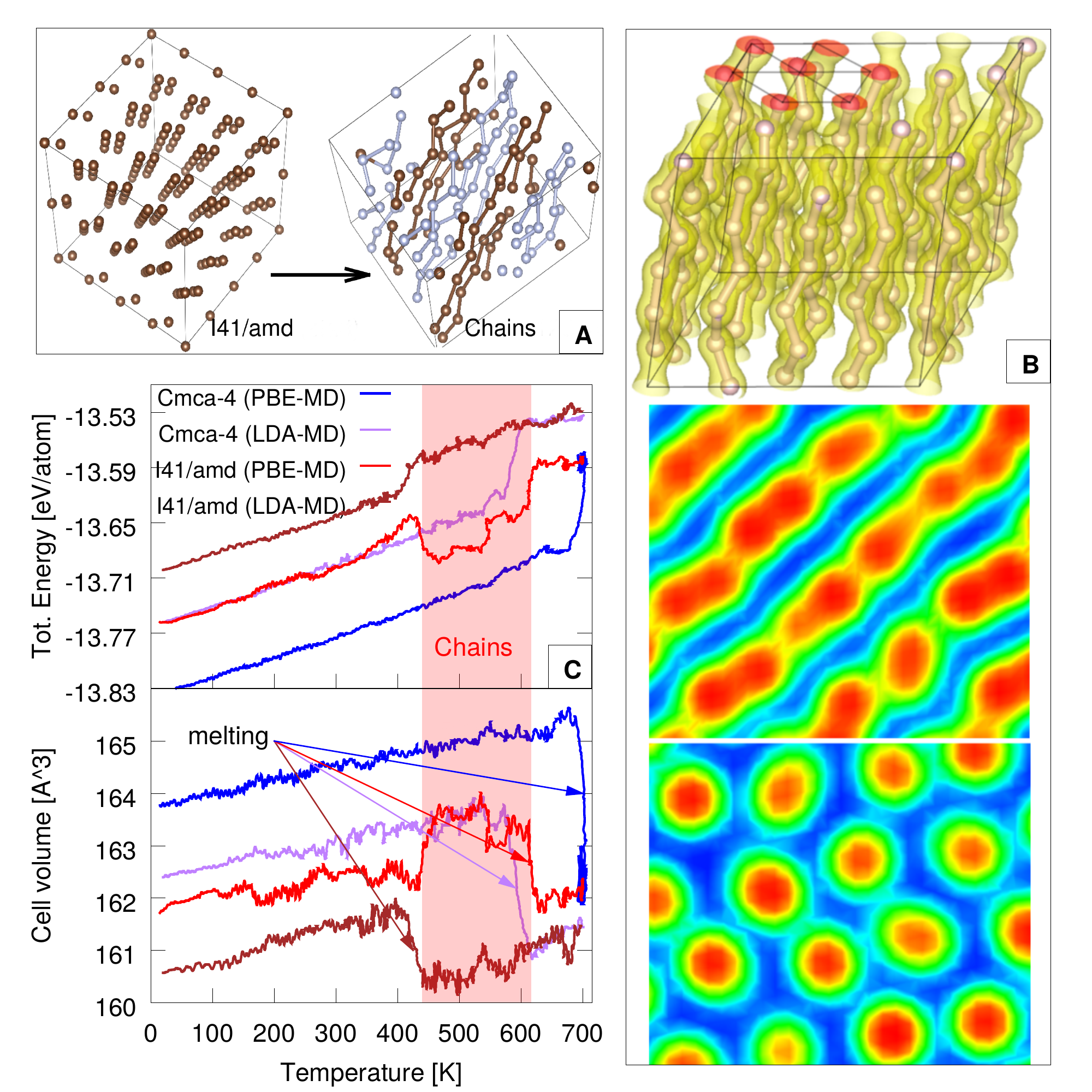}
\end{center}
\vspace{-12pt}
\caption{(A) Transformation of $I41/amd$ into chains
  wrapping through periodic boundaries. (B) MD snapshots showing the charge density iso-surface (top)
  and the ELF \cite{savin1992electron} (bottom) associated with $Chains$. (C) Cell volume and total energy
  upon heating at 400GPa (SM). Shaded in red is the region where $Chains$ appears.}
\label{chains}
\vspace{-20pt}
\end{wrapfigure}

Hydrogen is typically located atop group 1A in the periodic table, and
while the H$_2$ molecule has no analogue among the alkali metals, one
might expect similarities to appear if a non-molecular phase is
obtained at extreme conditions.

At pressures where the binding energy is close to zero, the structure
is held together by the external pressure rather than chemical
bonding. For the alkali metals in this pressure range, complex phases
are observed in which the electronic structure can be assigned to two
classes. The relatively simple electride structures involve the
electron being located in the interstices between atoms playing the
role of pseudo-anion
\cite{ma2009transparent,gatti2010sodium,marques2009potassium,marques2011crystal}. Alternately,
the atoms arrange themselves so as to perturb a free electron gas into
a charge-density wave, opening a gap (or pseudogap) 
at the Fermi
surface - the signature here being a clustering of diffraction peaks
at
2k$_F$\cite{Heine,marques2011crystal,arapanaSc,arapanaCa,ReedBa2000,degty,ackland2004origin}.

Static experiments using diamond anvil cells are now reaching
equivalent pressures in hydrogen, where the work done in compression
is greater than the chemical binding energy.  Measurements at these
pressures are restricted to spectroscopy and conductivity, so density
functional calculations are used to determine the structure.  Despite
the wide variety of methods used, including different treatment of the
quantised nuclear motion and different approaches to the
exchange-correlation functional, the same set of candidate structures
invariably appears, albeit with considerable variation in the
stability range.  Furthermore, more accurate theoretical treatments
invariably come at high computational cost, so at elevated temperature 
enhanced finite sampling errors can eliminate theoretical improvements.

At present there is no good consensus about the best trade-off:
typical calculations have errors of up to 100GPa and a few hundred
Kelvin.  Thus the practical role of calculations is to propose
candidate structures for test against experiment.

\section{Calculations}

\begin{wrapfigure}{lh}{0.6\textwidth}
\vspace{-30pt}
\begin{center}
\includegraphics[width=0.6\textwidth]{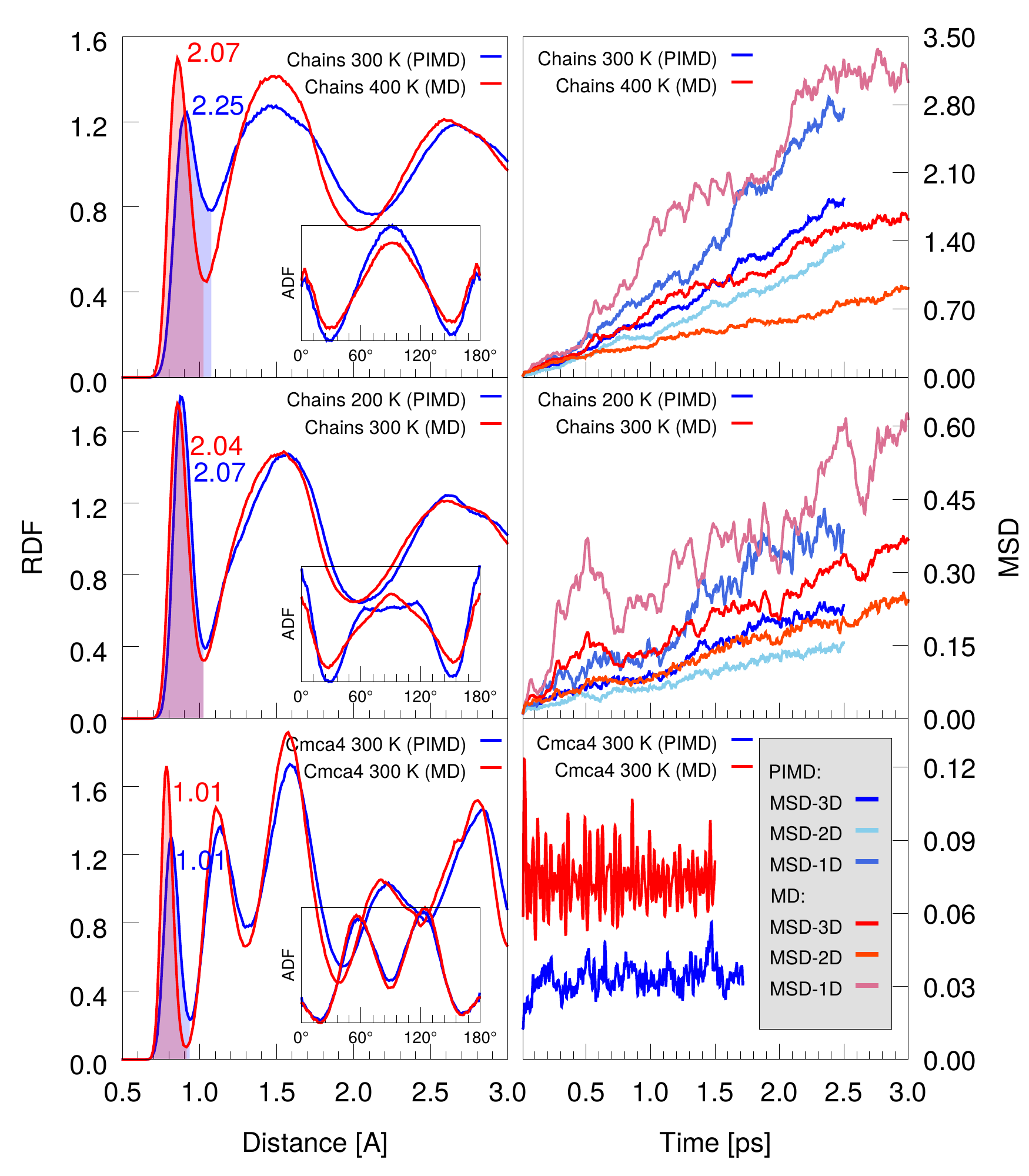}
\end{center}
\vspace{-25pt}
\caption{The left panels show RDFs and ADFs (insets)
  for $Chains$ and $Cmca-4$ calculated from both MD and PIMD at 400GPa and temperatures as labeled. Shaded is the integrated area
  under the first peak of RDF which gives the coordination number. The right panels depict the corresponding
  MSD calculated as total displacement (3D), displacement along chains (1D) and between chains (2D); normalised according to dimensionality.}
\label{rdf}
\vspace{-30pt}
\end{wrapfigure}
For analysis we use the radial distribution function (RDF), angular
distribution function (ADF), projected mean square displacement (MSD),
phonon and vibrational mode projection; details of which are given in
the SM. In addition we calculate average electron density of states (DoS)
and average XRD patterns, all explained in SM.

\section{Results and discussions}

We performed 17 short MD simulations starting with different
structures found using static structure
searches\cite{pickard2007structure,pickard2012density, geng2012high}.
Simulations starting in molecular phases retained molecular character,
but all high-symmetry structures are unstable and immediately undergo
total reconstruction at finite temperature.  Surprisingly, they adopt
rather similar structures with twofold coordinated chains of hydrogen
threading through the simulation box, subject to the constraints of
periodic boundary conditions (Figure \ref{chains}). This $Chains$
structure, has on average 2D hexagonal symmetry in the plane
perpendicular to the chains, while being essentially a collection of
charge tubes containing diffusive hydrogen nuclei in the chain
direction. Similar structures have been identified before in structure searches 
as transition states between the molecular and atomic phases\cite{geng2012high}
and probably the same structure was reported in previous PIMD
simulations\cite{Biermann, BiermannX}.  Its ubiquity marks it out as a
candidate high temperature structure worthy of further study.
\begin{wrapfigure}{rh}{0.6\textwidth}
\begin{center}
\vspace{-35pt}
\includegraphics[width=0.6\textwidth]{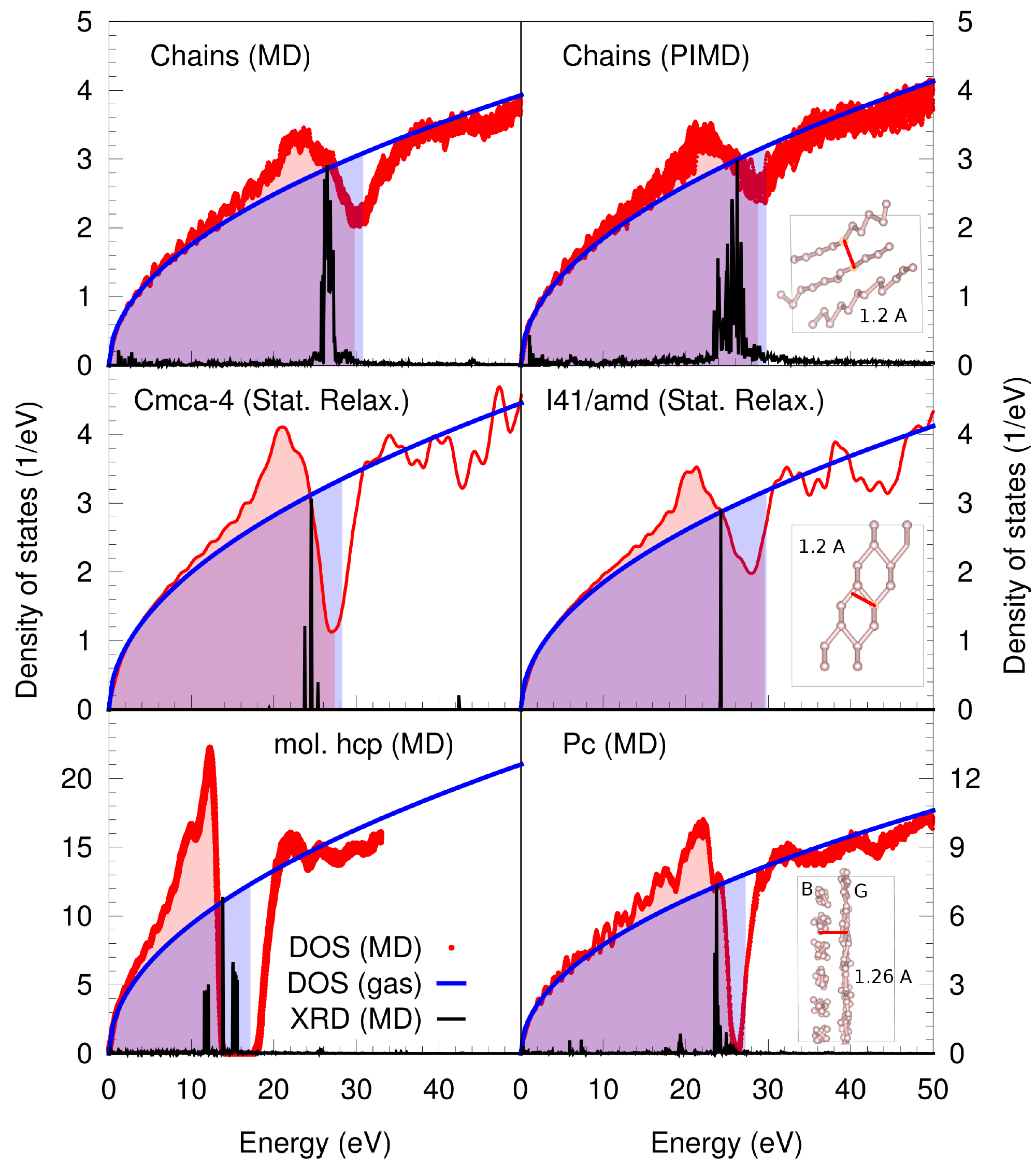}
\end{center}
\vspace{-15pt}
\caption{The figure shows in red the density of electronic states (DoS)
  from multiple snapshots in the MD/PIMD at 300K and from
  static relaxations (middle panels) for comparison. The pressure is different for each phase candidate:
  $mol.\:hcp$ (50GPa), $Pc$ (275GPa), $Cmca-4$, $I41-amd$ and $Chains$ (all at 400GPa). Blue lines are the
  analytic DoS for a free electron gas at the same density.
  Shaded regions depict the occupied states. In solid black
  we show the simulated XRD powder pattern (calculated with GDIS \cite{fleming2005gdis})
  averaged over the MD/PIMD snapshots. Only the relative intensities are meaningful, and the peak positions are
  plotted in units of energy of a free electron with equivalent wavelength. The insets show the length scale corresponding to the most intense peak.}
\label{bands}
\vspace{-15pt}
\end{wrapfigure}
In addition to these short runs, we performed two long MD simulations
using a 2x2x2 k-point grid at 400GPa, starting in $I41/amd$ (atomic)
and $Cmca-4$ (molecular) and slowly heated (Figure \ref{chains}).
Using PBE exchange-correlation, the $I41/amd$
spontaneously transformed into $Chains$ at around 450K which remained
stable until melting occurred at 620K. Although, $Chains$ is clearly
more thermodynamically stable than $I41/amd$, compared to $Cmca-4$,
which remains molecular until melting, $Chains$ has higher energy, but
also lower volume and probably higher entropy owing to large diffusion
(Figure \ref{rdf}).

Figure \ref{chains} also shows that the liquid is
denser than any candidate solid structure at 400GPa, confirming the
negative slope of the melting curve \cite{howie2015raman}. $Cmca-4$
would give a much steeper slope than $Chains$, which could be measured
experimentally. Using LDA exchange-correlation, both phases melt at
lower temperatures, indicating that choice of exchange-correlation
functional can change melting temperatures by around 100K
(corresponding to about 10meV).

We find that diffusion in the $Chains$ structure is significantly
higher than in the competing molecular metallic phases (Figure
\ref{rdf}). Studying the atomic trajectories in detail we identify 3
main mechanisms; in order of importance: a) chains sliding "freely"
along their length (included in MSD-1D), b) atoms migrating to
neighbour chains (MSD-2D) and c) atoms swapping position within the
chain (included in MSD-1D), sometimes by forming a short-lived
rotating molecule. The diffusive nature of $Chains$ makes it a higher
entropy structure which might stabilize it over $Cmca-4$ at higher
temperature.  For both $Chains$ and $Cmca-4$ the effect of quantum
protons (PIMD) on MSD, RDF and ADF is similar to increasing the
temperature by about 100K in classical MD.
The RDF confirms that $Chains$ has coordination number 2 with a peak at $0.9\:A$, meaning the molecules
are no longer well defined. We notice in the ADF a well defined peak at $90^{\circ}$ which means that the pseudo-molecules
are pointing towards one-another. $Cmca-4$ can be viewed as a similar structure, but with the chains broken into well defined
molecules, oriented at angles with respect to the chain direction. We can say $Chains$ is a high entropy version of $Cmca-4$.

Figure \ref{bands} shows the electronic density of states (DoS) taken
from individual snapshots of MD. DoS is extremely well fitted by a
free-electron gas model, with a pronounced dip at the Fermi energy.
To open a pseudogap at the Fermi surface of a free electron gas it is
necessary to have an arrangement of atoms which will strongly scatter
electrons at k$_F$.  Such an arrangement will also strongly scatter
X-rays at $q=2k_F$. In figure \ref{bands} we relate the X-ray
scattering vector to an equivalent energy using $E_q=\hbar^2q^2/8m$.
We have used GDIS software\cite{fleming2005gdis} to plot the simulated
X-ray diffraction pattern (XRD) implied by our simulations.  It can be
seen that the $Chains$ structure self-organizes in such a way that a
large number of symmetry-unrelated peaks are co-located at $E_q\lesssim
E_f$. 
\begin{wrapfigure}{lh}{0.6\textwidth}
\begin{center}
\vspace{-30pt}
\includegraphics[width=0.6\textwidth]{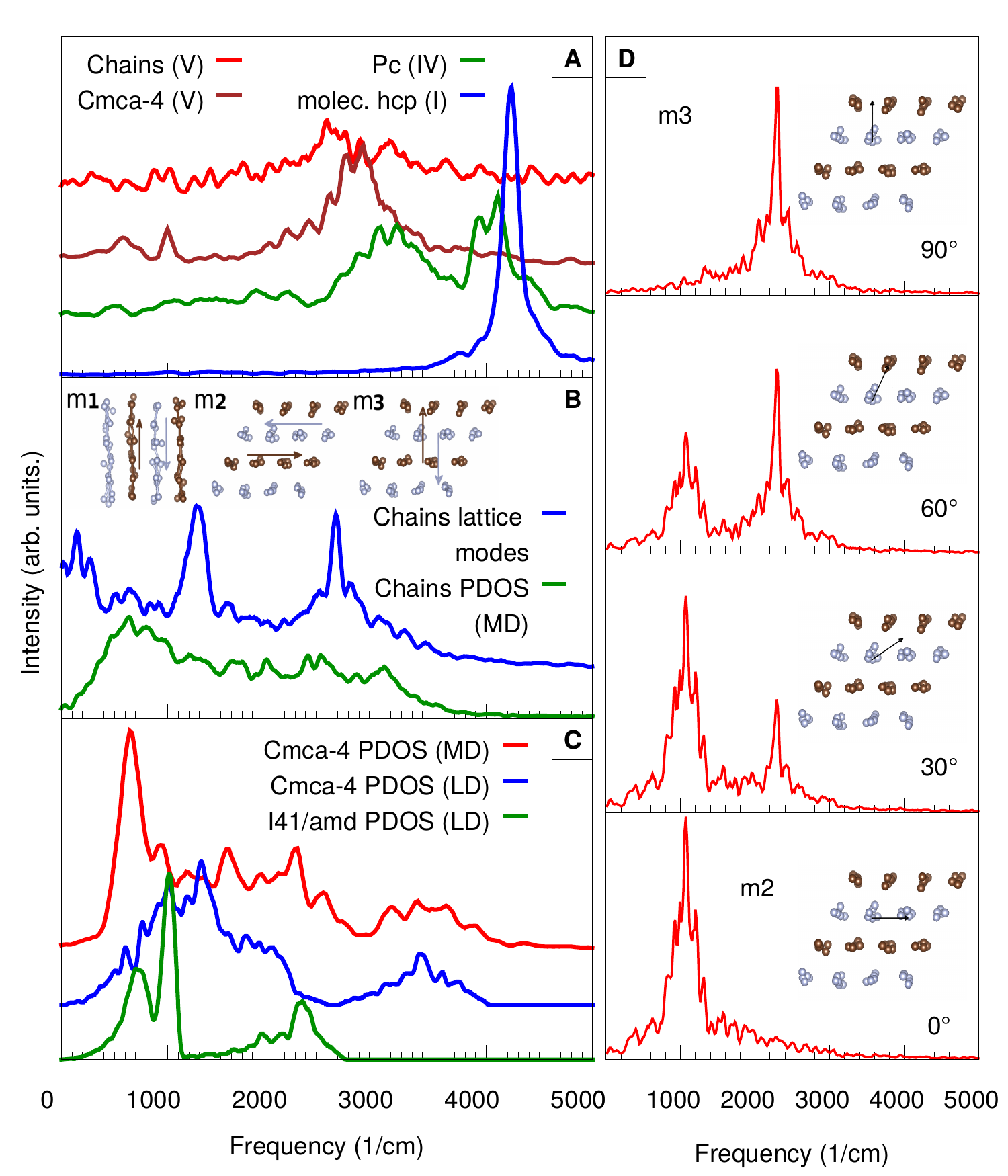}
\end{center}
\vspace{-20pt}
\caption{(A) Fourier transformed projection of the
  MD velocities onto molecular stretches
  \cite{magduau2013identification, ackland2013efficacious,
  magduau2013high} for the candidates of phases I, IV and
  beyond. (B) Phonon density of states (PDoS) extracted from 
  VACF together with projections onto the most symmetric modes of the $Chains$ (also depicted in the insets), all
  extracted from MD trajectories. (C) PDoS calculated from MD (300K) and LD (0K) for $Cmca-4$ and $I41/amd$.
(D) illustrates how the the most symmetric modes emerge as the eigenvector is rotated from $0^{\circ}$ to $90^{\circ}$ .}
\label{raman}
\vspace{-13pt}
\end{wrapfigure}
 This is identical to the situation in the alkali metals, in
particular for the incommensurate
phases\cite{RbChain,ackland2004origin} which typically have chains of
atoms in a ``host'' framework. It is thus clear that our $Chains$
structure can be interpreted as a Fermi-surface perturbation from a
free electron gas.  This nearly-free electron picture, stabilization
by a charge density wave, appears to be valid for other candidate
hydrogen structures.  In the case of Phase-IV candidate, $Pc$, the
charge density wave creates a layered structure (insets to figure
\ref{bands}) which can accommodate molecules and hexamers.  This could
explain the preferences of most structure candidates in the pressure
range close to metalization to arrange in layers or polymers.  It is
peculiar that even at 50GPa, the molecular structure
$mol. hcp.$ appears to have a nearly free electron DoS. This might be
a side effect of the plane wave approximation employed in DFT.

In figure \ref{raman} we show our full phonon spectrum (PDOS),
calculated from the fourier transform of the velocity autocorrelation
function for $Cmca-4$ and $Chains$ and from lattice dynamics (LD) for
$Cmca-4$ and $I41/amd$. The spectra obtained for $Cmca-4$ at 0K and
300K are consistent suggesting the phonons are harmonic.  While
$I41/amd$ presents three distinct bands in the PDOS at 0K, the
$Chains$ structure has only one broad band of frequencies with no gap.

Because $Chains$ is metalic and only stable in MD, there is no way to identify the Raman modes,
we can only impose restrictions based on symmetry. We therefore calculate the frequencies for
the most symmetric phonon modes, involving oscillations of chains, by projecting the MD velocities onto
those motions, and taking the FT of the resulting trajectory (see SM for details). There are three
main modes involving chain oscillations in the 3 Cartesian directions as depicted in figure \ref{raman}. 
We also demonstrate that our method correctly picks up the eigen-frequencies by investigating the
spectrum with respect to different projection vectors.

Since the $Chains$ have no molecular modes, so 
there are no vibronic modes above 3000cm$^{-1}$ in the PDOS.  The
highest frequency modes in the calculation are around 2500cm$^{-1}$,
which comprise atoms vibrating within a chain.  Occasionally, we
do see interatomic distances comparable with the free molecule, 
but these are short lived. $Chains$ is a natural continuation of the molecular phases
on the path to dissociation, as illustrated by the vibrational modes (Figure \ref{raman})
given by the different structure candidates in order of increasing pressure.

\section{Simulation shortcomings}
\begin{wrapfigure}{rh}{0.59\textwidth}
\vspace{-55pt}
\begin{center}
\includegraphics[width=0.59\textwidth]{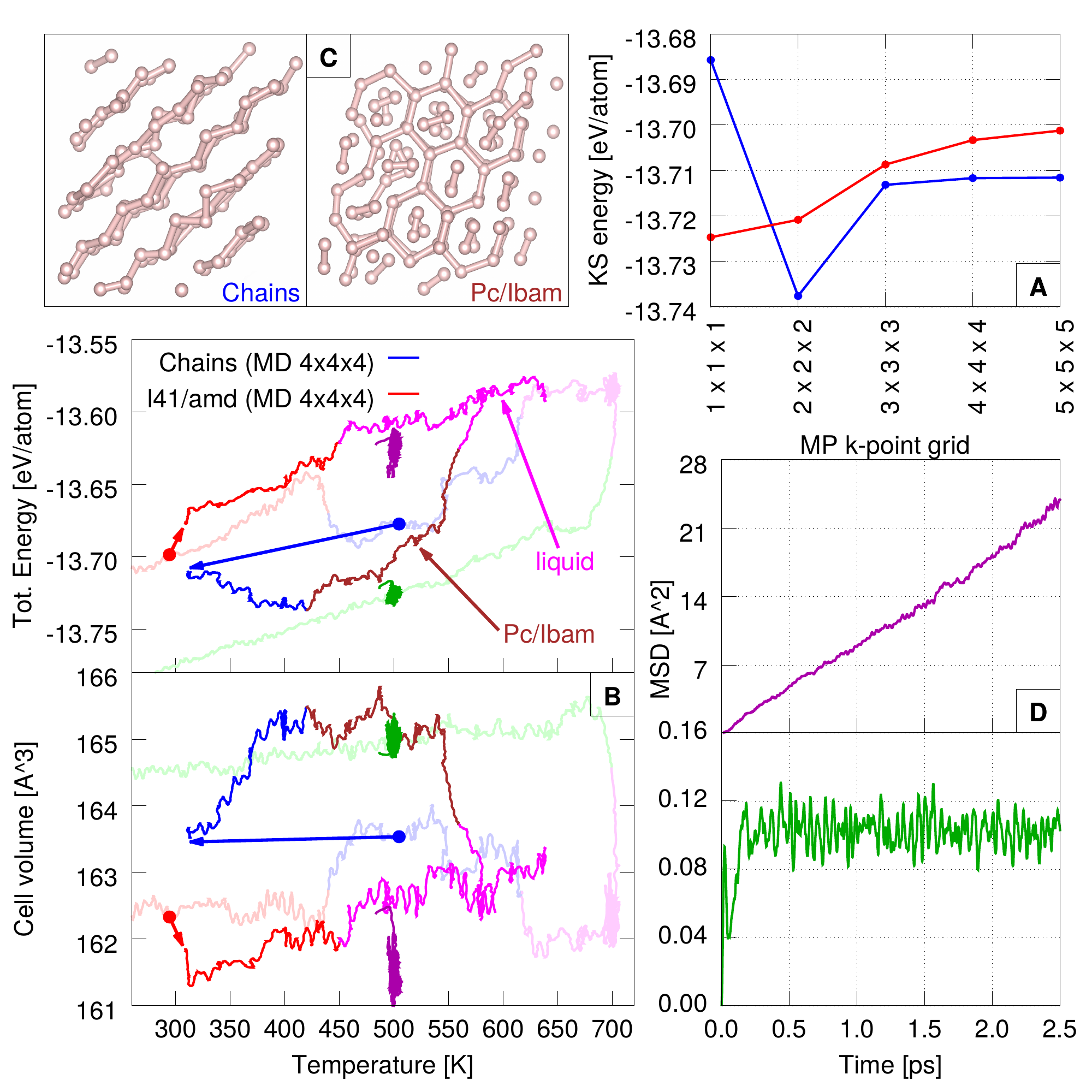}
\end{center}
\vspace{-22pt}
\caption{(A) Single point calculations of energy convergence
 with k-point sampling for $Chains$ and $I41/amd$. (B) Heating at 400GPa with 
increased k-points (4x4x4), starting from snapshots of (2x2x2) MD runs (Figure \ref{chains}).
 The colors code the different structures: $Cmca-4$ (green), $Chains$ (blue),
 $Pc/Ibam$ (brown), $liquid$ (purple). (C) Transitions from 
$Chains$ to $Pc/Ibam$ through snapshots from previous MD runs. (D) MSD of MD simulations started in $I41/amd$ and $Cmca-4$, at 400GPa and 
500K with 2x2x2 k-points.}
\label{kpoint}
\vspace{-15pt}
\end{wrapfigure}
The solid phases of hydrogen discovered so far at room temperature are stabilised by
 entropy, therefore MD simulations are crucial for understanding the properties of
 possible candidates. These calculations
 are expensive, however, and we need to consider trade-offs.

A large size of the cell is important to reduce finite size effects, a
lengthy run is needed for equilibration and good sampling of the
energy surface.  At the same time, a reasonably small time step is
required to capture the hydrogen molecular vibrations. In addition,
for PIMD runs, a large number of beads is necessary to sample the
quantum wave function of the nuclei. These are all supplementary to
the actual DFT settings that mainly include the choice of functional,
electronic energy cut-off and k-point sampling. The accuracy of the
PBE functional for hydrogen has vastly been disputed in the
literature, however, we emphasize that in the case of MD using PBE is
a reasonable trade-off to minimize the other more important
shortcomings.

We draw attention here to yet another convergence issue that has not
been fully investigated before. It is common practice to run large MD
calculations with small k-point sets, as appropriate for the large
cell size.  In MD runs with a single k-point, we noticed that nuclei
can rearrange in real space such that the energy of the electrons is
minimized at $\Gamma$, but not at unsampled points in the zone.  We
demonstrate the effect using calculations (shown in Figure \ref
{kpoint}A) on snapshots taken from our previously discussed MD heating
runs (Figure \ref{chains}) comparing the pre-transition $I41/amd$
structure with the post-transition $Chains$.  We see that using
$\Gamma$-point only sampling, $I41/amd$ has the lowest energy, whereas
in the actual simulation, using a 2x2x2 grid, $Chains$ is lower.  For
higher K-point sampling, the $Chains$ snapshots are still lower in
energy than $I41/amd$, but by a smaller amount.  This suggests that
although the $Chains$ configuration may be optimised to match the
$k$-point grid, it is still a valid candidate structure.  Previous MD
calculations have, at best, investigated $k$-point convergence with
respect to static structures and often details of $k$-points are not
even quoted.  The 20meV uncertainty suggests that $k$-point sampling,
rather than functional or quantum-proton effects, may be the dominant
source of error in previous MD calculations.

Figure \ref{kpoint} also compares the
previous MD runs (Figure \ref{chains}) with a subsequent MD heating
simulation with a denser MP grid, in which $I41/amd$ melts at 450K
and $Chains$ transforms to a $Pc/Ibam$ type structure. In constant
pressure and temperature simulations at 400GPa and 500K, $I41/amd$
melts whereas $Cmca-4$ is stable. This new results suggest that the
stability ranking at these conditions is: $Cmca-4$, $Pc/Ibam$,
$Chains$, $I41/amd$. Of course, we must study all competitive
candidates, leaving the experiment to decide.

\section{Conclusions}
We have found and investigated a candidate structure for
high-temperature solid hydrogen under pressure. The polymeric $Chains$
structure has twofold coordinated hydrogen atom, and is stabilized by
a charge density wave similar to the incommensurate structures in the
simple metals.  In experiment, it would be characterized by the
disappearance of high frequency vibronic peak.  Up to 400GPa, $Chains$
is clearly more stable than any network atomic structure, but its
stability relative to molecular structures such as $Cmca-4$ cannot be
reliably determined using our methods.

\section{Acknowledgments}
We thank E.Gregoryanz, A.Hermann and
  M.Martinez-Canales for many useful discussions. IBM thanks the Cray XC30 for
  computing time (Archer at EPCC) and EPSRC for a studentship.

\section*{References}

\bibliographystyle{iopart-num}
\bibliography{Refs}

\providecommand{\newblock}{}
\begin{thebibliography}{10}
\expandafter\ifx\csname url\endcsname\relax
  \def\url#1{{\tt #1}}\fi
\expandafter\ifx\csname urlprefix\endcsname\relax\def\urlprefix{URL }\fi
\providecommand{\eprint}[2][]{\url{#2}}

\bibitem{savin1992electron}
Savin A, Jepsen O, Flad J, Andersen O~K, Preuss H and von Schnering H~G 1992
  {\em Angewandte Chemie International Edition in English\/} {\bf 31} 187--188

\bibitem{ma2009transparent}
Ma Y, Eremets M, Oganov A~R, Xie Y, Trojan I, Medvedev S, Lyakhov A~O, Valle M
  and Prakapenka V 2009 {\em Nature\/} {\bf 458} 182--185

\bibitem{gatti2010sodium}
Gatti M, Tokatly I~V and Rubio A 2010 {\em Physical Review Letters\/} {\bf 104}
  216404

\bibitem{marques2009potassium}
Marqu{\'e}s M, Ackland G~J, Lundegaard L~F, Stinton G, Nelmes R~J, McMahon M~I
  and Contreras-Garcia J 2009 {\em Physical Review Letters\/} {\bf 103} 115501

\bibitem{marques2011crystal}
Marqu{\'e}s M, McMahon M~I, Gregoryanz E, Hanfland M, Guillaume C~L, Pickard
  C~J, Ackland G~J and Nelmes R~J 2011 {\em Physical Review Letters\/} {\bf
  106} 095502

\bibitem{Heine}
Heine V 2000 {\em Nature\/} {\bf 403} 836–837

\bibitem{arapanaSc}
Arapan S, Skorodumova N~V and Ahuja R 2009 {\em Physical Review Letters\/} {\bf
  102} 085701

\bibitem{arapanaCa}
Arapan S, Mao H and Ahuja R 2008 {\em PNAS\/} {\bf 105} 20627

\bibitem{ReedBa2000}
Reed S~K and Ackland G~J 2000 {\em Physical Review Letters\/} {\bf 84} 5580

\bibitem{degty}
Degtyareva V~F 2006 {\em Phys. Uspek.\/} {\bf 49} 369–388

\bibitem{ackland2004origin}
Ackland G~J and Macleod I~R 2004 {\em New Journal of Physics\/} {\bf 6} 138

\bibitem{pickard2007structure}
Pickard C~J and Needs R~J 2007 {\em Nature Physics\/} {\bf 3} 473--476

\bibitem{pickard2012density}
Pickard C~J, Martinez-Canales M and Needs R~J 2012 {\em Physical Review B\/}
  {\bf 85} 214114

\bibitem{geng2012high}
Geng H~Y, Song H~X, Li J and Wu Q 2012 {\em Journal of Applied Physics\/} {\bf
  111} 063510

\bibitem{Biermann}
Biermann S, Hohl D and Marx D 1998 {\em Solid State Communications\/} {\bf 108}
  337--341

\bibitem{BiermannX}
Biermann S, Hohl D and Marx D 1998 {\em J. Low Temp. Phys.\/} {\bf 110} 97--102

\bibitem{fleming2005gdis}
Fleming S and Rohl A 2005 {\em Zeitschrift Fur Kristallographie\/} {\bf 220}
  580--584

\bibitem{howie2015raman}
Howie R~T, Dalladay-Simpson P and Gregoryanz E 2015 {\em Nature materials\/}
  {\bf 14} 495--499

\bibitem{magduau2013identification}
Magd{\u{a}}u I~B and Ackland G~J 2013 {\em Physical Review B\/} {\bf 87} 174110

\bibitem{ackland2013efficacious}
Ackland G~J and Magdau I~B 2013 {\em High Pressure Research\/}  1--7

\bibitem{magduau2013high}
Magd{\u{a}}u I~B and Ackland G~J 2014 {\em J. Phys.: Conf. Ser.\/} {\bf 500}
  032012

\bibitem{RbChain}
McMahon M~I and Nelmes R~J 2004 {\em Physical Review Letters\/} {\bf 93} 055501

\end{thebibliography}

\end{document}